\documentclass[letterpaper]{article} 
\usepackage{aaai2026}  
\usepackage{times}  
\usepackage{helvet}  
\usepackage{courier}  
\usepackage[hyphens]{url}  
\usepackage{graphicx} 
\usepackage{natbib}  
\usepackage{caption} 
\newcommand{\tool}{\textsc{Steamroller}}

\usepackage{multirow}
\usepackage{booktabs}
\usepackage{algorithm}
\usepackage{algpseudocode}
\usepackage{amsmath}
\usepackage{amsfonts}

\makeatletter  
\newif\if@restonecol  

\usepackage{newfloat}
\usepackage{listings}
\DeclareCaptionStyle{ruled}{labelfont=normalfont,labelsep=colon,strut=off} 
\floatstyle{ruled}
\newfloat{listing}{tb}{lst}{}
\floatname{listing}{Listing}
\gdef\copyright@on{}      

\title{\tool{}: A Multi-Agent System for Inclusive Automatic Speech Recognition for People Who Stutter}

\author {
    Ziqi Xu\textsuperscript{\rm 1},
    Yi Liu\textsuperscript{\rm 2},
    Yuekang Li\textsuperscript{\rm 3},
    Ling Shi\textsuperscript{\rm 4},
    Kailong Wang\textsuperscript{\rm 5},
    Yongxin Zhao\textsuperscript{\rm 1}\thanks{Corresponding author.}
}
\affiliations {
    \textsuperscript{\rm 1}Shanghai Key Laboratory of Trustworthy Computing, East China Normal University, Shanghai, China\\
    \textsuperscript{\rm 2}AI Research Team, Quantstamp\\
    \textsuperscript{\rm 3}The University of New South Wales, Australia\\
    \textsuperscript{\rm 4}Nanyang Technological University, Singapore\\
    \textsuperscript{\rm 5}Huazhong University of Science and Technology, China\\
    51265902070@stu.ecnu.edu.cn, yi009@e.ntu.edu.sg, yuekang.li@unsw.edu.au, ling.shi@ntu.edu.sg,\\
    wangkl@hust.edu.cn, yxzhao@sei.ecnu.edu.cn
}

\begin{document}

\maketitle

\begin{abstract}
People who stutter (PWS) face systemic exclusion in today’s voice-driven society, where access to voice assistants, authentication systems, and remote work tools increasingly depends on fluent speech. Current automatic speech recognition (ASR) systems, trained predominantly on fluent speech, fail to serve millions of PWS worldwide. We present \tool{}, a real time system that transforms stuttered speech into fluent output through a novel multi-stage, multi-agent AI pipeline. Our approach addresses three critical technical challenges: (1) the difficulty of direct speech to speech conversion for disfluent input, (2) semantic distortions introduced during ASR transcription of stuttered speech, and (3) latency constraints for real time communication. \tool{} employs a three stage architecture comprising ASR transcription, multi-agent text repair, and speech synthesis, where our core innovation lies in a collaborative multi-agent framework that iteratively refines transcripts while preserving semantic intent.  Experiments on the FluencyBank dataset and a user study demonstrates clear word error rate (WER) reduction and strong user satisfaction. Beyond immediate accessibility benefits, fine tuning ASR on \tool{} repaired speech further yields additional WER improvements, creating a pathway toward inclusive AI ecosystems.
\end{abstract}

\section{Introduction}
Speech disorders affect 90-95\% of the global population at various stages~\cite{bloodstein2021handbook}, with stuttering being a prevalent neurodevelopmental disorder characterized by involuntary disruptions including blocks, prolongations, and repetitions. For people who stutter (PWS), these disruptions create cascading societal impacts: students may avoid classroom participation, professionals miss career opportunities requiring presentations or client interactions, and individuals withdraw from social situations to avoid stigma and misunderstanding. In our increasingly voice driven society, PWS face compounding discrimination as they cannot reliably use voice assistants for basic tasks, are excluded from voice authenticated services, and find themselves disadvantaged in remote work environments that rely heavily on video calls. Traditional interventions such as speech therapy and electronic fluency devices, while beneficial for some, often require years of commitment with variable outcomes and fail to address the immediate technological and social barriers that systematically marginalize PWS in education, employment, and civic participation.

Despite advances in automatic speech recognition (ASR) through deep learning architectures~\cite{6638947,7953077,Hannun2014DeepSS,wav2vec}, current systems deployed by major technology companies~\cite{google-asr,azure-asr,ibm-asr} systematically fail to accommodate stuttered speech. These systems, designed and tested primarily on fluent speech, result in recognition failures and exclusion from voice based interfaces~\cite{bleakley2022exploring,Clark,StammerApp}. While generative AI presents opportunities for real time speech processing that could potentially transcribe, repair disfluencies, and synthesize corrected speech, these capabilities remain largely unexplored for accessibility applications. This technological gap highlights the urgent need for inclusive ASR design that recognizes stuttering not as an edge case but as a critical accessibility requirement affecting millions worldwide.

To address this societal and accessibility gap, we explore the development of a system that transforms stuttered audio into fluent, clear speech. Crucially, achieving these goals requires addressing three challenges:  \textbf{C\#1}: Modifying speech at the content level remains highly challenging. In particular, converting streaming stuttered audio directly into fluent speech is difficult given the current limitations of speech modeling technologies. \textbf{C\#2}: Semantic loss may occur during the transcription process~\cite{Lea2021SEP28kAD}. It is crucial to preserve the semantically correct portions of the original speech while selectively repairing segments that may contain semantic errors, ensuring that the final output accurately and faithfully conveys PWS's intended message. \textbf{C\#3}: Since the system is designed for real-time communication, the entire repair process must be efficiently managed to maintain low latency.

In this work, we introduce \tool{}, which adopts a three-stage pipeline to address the challenges: transcribing stuttered audio, repairing the text, and synthesizing fluent speech. Instead of direct conversion, we use an intermediate textual representation to improve controllability and reliability. For repair, we design a multi-agent system that refines transcription with semantic consistency. To support real-time use, we minimize latency in each stage without sacrificing quality.

We first conducted experiments on the FluencyBank~\cite{fluency-bank} dataset to evaluate repair effectiveness. Across ASR systems, \tool{} achieves 10.46\% WER, 8.57\% MER, and 11.40\% WIL reduction. It also improves semantic similarity by 10\% over original ASR transcriptions. \tool{} enhances FluencyBank with stutter-free references (validated by the dataset authors) and improves annotation quality for SEP-28K~\cite{Lea2021SEP28kAD}. Fine-tuning ASR with repaired speech yields an additional 3\% WER drop. We also conducted a 23-participant user study. In the close-ended task study, \tool{} reduces WER by 7–11\% around for moderate/severe stuttering and 3\% for mild cases. Participants highly recognized its repair effectiveness, with a satisfaction score of $4.30 \pm 0.27$ ($p \ll 0.01$). The open-ended task study investigates real-time use with PWS, showing positive user feedback (NPS = 23.53, SUS = 66.9). Overall, our approach effectively repairs stuttering in speech, highlighting its potential as a real-time stuttering audio translator that enhances communication. These applications illustrate the versatility and impact of \tool{} in real-world scenarios.

This paper makes the following contributions to the field of AI for Social Impact:
\begin{itemize}
    
    \item We formalize the challenge of repairing semantically distorted ASR transcripts for PWS and justify our multi-stage, multi-agent pipeline design through systematic trade-off analysis.
    \item We present \tool{}, a system which combines ASR, LLMs, and speech synthesis via a novel multi-agent framework. It models collaborative reflection to resolve semantic inconsistencies in disfluent speech.
    \item We conduct mixed-methods evaluation with 23 PWS participants, combining quantitative FluencyBank benchmark analysis~\cite{fluency-bank} with qualitative feedback from conversational tasks.
    \item We demonstrate dual utility as both an assistive tool for PWS and a framework for dataset enhancement, showing that fine-tuning ASR on our system's output improves accuracy and enables more inclusive AI development.
\end{itemize}

\section{Background and Related Work}

\subsection{Stuttering and Speech Disorders}

Stuttering is a multifaceted fluency disorder that disrupts speech flow and hampers communication~\cite{asha,laiho2022stuttering}. Its origins remain unclear, involving both biological and psychological factors~\cite{tichenor2019stuttering}, often leading to social isolation and emotional distress~\cite{craig2014trait}. Speech therapy aims to improve fluency, reduce anxiety, and boost self-esteem~\cite{murza2019effects, brignell2020systematic}. Stuttering is categorized into five types-blocks, prolongations, sound repetitions, word/phrase repetitions, and interjections~\cite{Lea2021SEP28kAD}, each posing distinct challenges for ASR recognition. This indicates the persistent difficulties PWS face when using ASR in modern smart devices~\cite{bleakley2022exploring,Lea_Colin}. Existing study~\cite{Lea_Colin} suggests that improved speech technology accuracy could encourage more frequent use by PWS. This gap motivates our investigation into ASR system assessment using generated stuttering audio. By doing so, we aim to improve ASR precision and promote ASR accessibility for PWS.

\subsection{ASR System Accessibility Enhancement}

The application of machine learning and deep learning to enhance ASR for PWS is gaining increasing attention. Recent studies have reviewed stuttering classification methods~\cite{asre-1}, evaluated machine learning techniques for stuttering event detection~\cite{asre-2,Sharma}, and explored multi-task and adversarial learning for improved feature extraction~\cite{asre-3}. Deep neural networks have also advanced stutter detection~\cite{asre-4}, while efforts continue to identify disfluency boundaries in children's speech~\cite{asre-5}.

Most existing efforts focus on detecting or labeling disfluent speech, rather than improving ASR system output for PWS. However, stuttered speech can lead to semantic-level distortions in ASR transcripts, which are often overlooked in prior work. To address this gap, we propose a novel multi-agent framework powered by LLMs that not only identifies and processes disfluencies but also actively restores semantic fidelity, significantly enhancing the usability of ASR systems for PWS.

\begin{figure}[t]
    \centering
    \includegraphics[width=\linewidth]{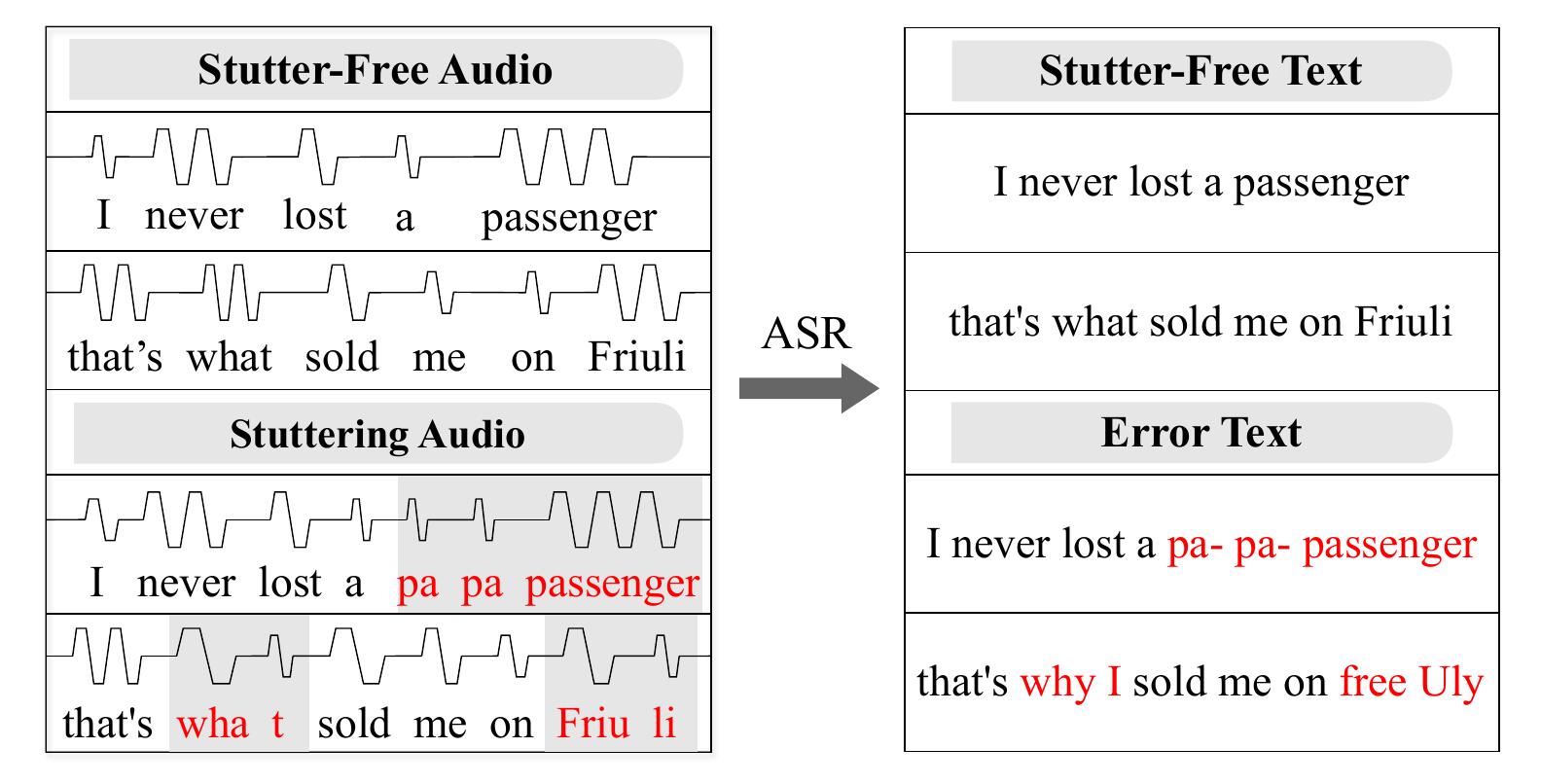}
    \caption{Comparison of ASR transcription outcomes for stutter-free and stuttering audio inputs. For the same sentence, stuttering speech may lead to errors in recognizing words and ultimately transcribing incorrect text, highlighting the inaccuracies in text generation from stuttered speech.}
    \label{motivation_example}
\end{figure}

\begin{figure*}[t]
    \centering
    \includegraphics[width=0.9\textwidth]{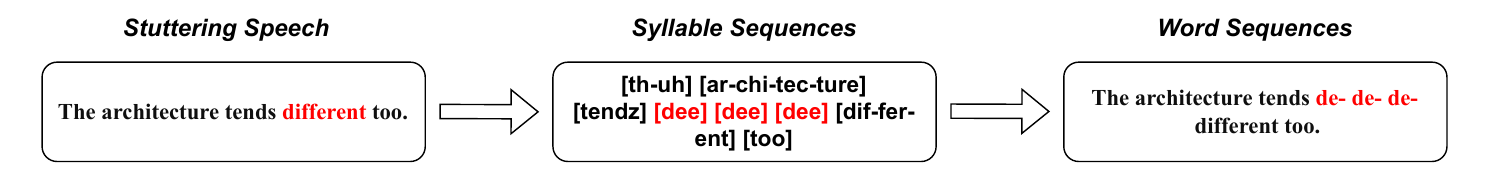}
    \caption{The process of stuttering speech to text conversion.  In this example, the word ``different'' includes several repeated syllables due to stuttering, which ultimately causes it to be incorrectly recognized as multiple words.}
    \label{ASR repair process}
\end{figure*}                                                                                       
\section{User Needs and Design Rationale}
\subsection{User Needs Analysis in \tool{}}

To identify user needs, we conducted 30-minute semi-structured online interviews with 16 PWS recruited from stuttering support communities. Discussions covered stuttering experiences, communication challenges, and expectations for ASR repair systems. Data were analyzed using Braun and Clarke’s Thematic Analysis~\cite{Braun01012006}, yielding four core needs:

\textbf{N1: Enhanced Speech Recognition.} One-fourth of participants reported that existing ASR systems misinterpret their disfluent speech. PWS require specialized recognition technology that accurately transcribes their speech patterns as a foundation for repair.

\textbf{N2: Stuttering Content Repair.} 
Two-thirds of participants noted their stuttering hindered listener comprehension, especially in professional settings. They need automatic detection and repair that produces natural, fluent output while reducing conversational stress.

\textbf{N3: Providing Pre- and Post-Repair Comparison.} 
One-third of participants wanted to see comparisons between original and repaired speech/text to evaluate system effectiveness and understand the repair process.

\textbf{N4: Real-Time Repair.} Two-thirds of participants emphasized the critical need for instant correction in formal settings like presentations, phone calls, and interviews where immediate, accurate speech is essential for effective communication.

\subsection{Design Rationale}

To address N1, \tool{} leverages GPT-4o to refine ASR transcriptions using task-specific prompts, enhancing accuracy and better capturing intended speech. 
For N2, \tool{} integrates the zero-shot voice cloning model StyleTTS2~\cite{common-voice} to synthesize fluent, stutter-free speech while preserving the speaker’s identity. This decoupled recognition-synthesis design improves communication accessibility and supports real-time use (N4), with short dialogues processed in approximately 3 seconds. 

Beyond this, voice cloning aids PWS in public speaking and enables clearer interaction with voice assistants, which often misinterpret disfluencies. The cloned speech can also augment ASR training data for improved recognition. 
To address N3, \tool{} offers a side-by-side comparison of original and repaired text/audio. For example, as shown in Figure~\ref{motivation_example}, an input like “I never lost a pa- pa- passenger” is repaired to “I never lost a passenger,” allowing users to intuitively assess correction quality.

\section{Design and Implementation}
\label{sec:methdology}
\begin{figure*}[t]
    \centering
    \includegraphics[width=\linewidth]{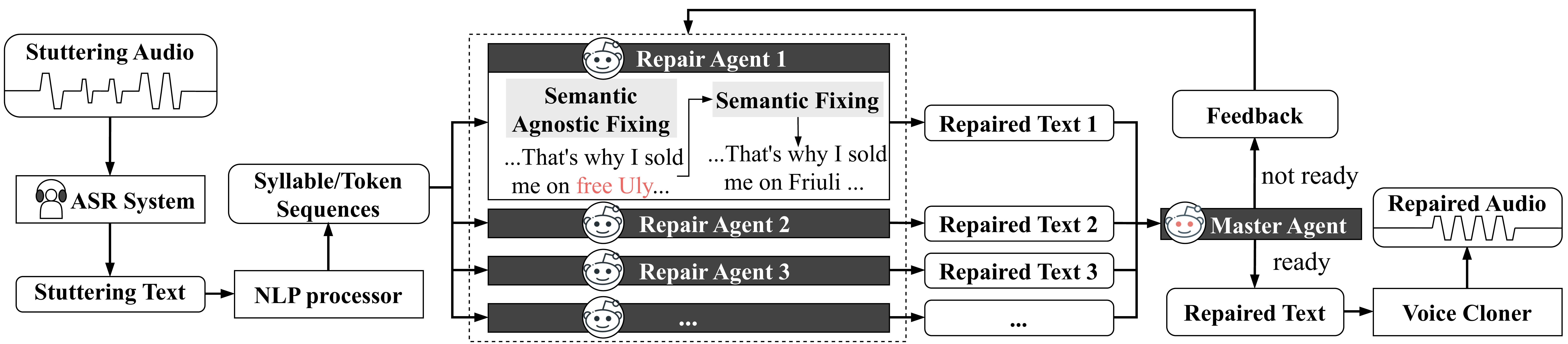}
    \caption{The workflow of \tool{}.  First, an ASR system transcribes the stuttered audio into text. Next, an NLP processor converts it into syllable or token sequences. Then, a multiagent cluster addresses potential misunderstandings in the transcribed text, with multiple repair agents responsible for making corrections. A master agent subsequently evaluates them and provides the final Repaired text. Finally, voice cloning technology transforms corrected text into stutter-free audio.}
    \label{asr_framework}
\end{figure*}

\tool{} employs a divide-and-conquer 
approach \cite{bentley1980multidimensional,smith1985design}, segmenting intricate issues into more manageable tasks. Direct conversion of stuttering audio to fluent audio while preserving semantic integrity poses significant challenges. Therefore, \tool{} focuses on two primary objectives: \textbf{Problem 1}, deciphering the intended content of stuttering audio; and \textbf{Problem 2}, converting this content into stutter-free audio without altering the speaker's unique voice characteristics. As shown in Figure~\ref{asr_framework}, our structured solution consists of three clear steps.

\subsection{Speech To Text Conversion}
\label{sec:speech-to-text-conversion}

To address \textbf{Problem 1}, \tool{} initially converts the stuttering audio into a textual transcript. The rationale is that established ASR systems, trained on extensive audio datasets, can capture a preliminary representation of the stuttering audio for subsequent refinement. To execute this step, as shown in Figure~\ref{ASR repair process}, \tool{} first converts stuttered audio $s_t$ into text $X = [x_1, x_2, ..., x_n]$ using an established ASR model such as Whisper~\cite{radford2023robust}. This step leverages ASR's pretrained capabilities to produce an initial transcription for further correction. 

However, as illustrated in Figure~\ref{ASR repair process}, stuttering often causes ASR models to misinterpret repeated syllables as valid words. These disfluent tokens, denoted $x_{i,p}$, are treated as legitimate inputs by the acoustic and language models, leading to semantic distortion in the transcript. resulting in $X=[x_1, x_2,..., x_{i,p}, x_{i+1,p},..., x_n]$. Despite the presence of stuttered content, language models maintain semantic coherence in sentence construction, providing a contextual basis for subsequent corrections.

\subsection{Semantic-aware Text Repairing}\label{sec:semantic-aware-text-Repairing}

Upon receiving ASR-transcribed text, \tool{} must correct inaccuracies caused by stuttering. This involves two key aspects: (1) systematically repairing different stuttering types without ground truth, and (2) defining termination criteria for the repair process in the absence of ground truth to prevent unnecessary iterations. 

\textbf{Design Rationale of Text Repairing.} The text repair module employs LLMs such as GPT-4o for iterative correction of ASR-generated text. Inspired by multi-agent collaboration, we design a coordinated cluster of AI agents to enhance correction stability and mitigate the inherent limitations of large models~\cite{kim2021conditional,Rewriting}. The cluster consists of two roles: a Repair Agent, which refines stuttering segments individually, and a Master Agent, which reviews the corrections and determines the termination point of the repair process.

To address misinterpreted text, our Repair Agent applies LLMs using prompt engineering methods such as Chain-of-Thought and Few-Shots Learning. This strategy is tailored to correct five distinct stuttering types, enhancing text accuracy by resolving stutter-induced inaccuracies effectively.
To determine the endpoint of the repair process and offer feedback on amended text, we deploy the Master Agent. This agent is responsible for deciding the conclusion of the repairing process and providing evaluative comments on the revised text to the Repair Agent for further refinement.

\begin{algorithm}[t]
\caption{Coordination Between Agents}
\label{algo:coordination}
\begin{algorithmic}[1]
\Require $S, prompt$
\Ensure $S'_{best}$
\State $S' \gets []$
\State Initialize $repair\_agents$
\State $Ready \gets \text{FALSE}$

\While{$\neg Ready$}
    \ForAll{$agent \in repair\_agents$}
        \State $S'.\text{append}(agent.\text{repair\_text}(S, prompt))$
    \EndFor
    \State $masterResult \gets MasterAgent(S')$
    \If{$masterResult.evaluation$}
        \State $Ready \gets \text{TRUE}$
    \Else
        \State $S \gets S'$
        \State $prompt \gets masterResult.feedback()$
    \EndIf
\EndWhile

\State \Return $masterResult.S'_{best}$
\end{algorithmic}
\end{algorithm}

\textbf{Coordination between Agents.} Algorithm~\ref{algo:coordination} outlines the procedural workflow of \tool{}, where the system initiates by deploying $n$ Repair Agents, each tasked with individually addressing inaccuracies in misinterpreted text segments. These Repair Agents operate independently to ensure a diverse range of potential corrections, enhancing the likelihood of identifying optimal solutions.

Once the initial repair attempts are completed, the Master Agent steps in to evaluate the progress based on three predetermined criteria: the stability of text improvements across iterations, the textual coherence with the original audio context, and a consensus among Repair Agents regarding the adequacy of text repairs. If the Master Agent decides that the repair objectives have been satisfactorily achieved, the process terminates, and the most recent set of repaired texts is finalized. However, if the criteria are not met-indicating potential-the Master Agent provides specific, constructive feedback to each Repair Agent. This feedback is based on discrepancies identified between the intended message and the current repaired text, guiding the agents towards more accurate modifications in the subsequent iteration. 

Based on empirical data gathered from processing over 95\% of audio files, we configure the system with three Repair Agents and a maximum of three iterations achieves a favorable trade-off between efficiency and repair quality.

\subsection{Text To Speech Conversion} \label{sec:text-to-speech-conversion}

To address \textbf{Problem 2}, \tool{} converts repaired text into fluent speech while preserving the speaker’s original voice. It uses a dual-encoder voice cloning approach: the Speaker Encoder captures vocal traits (e.g. pitch, timbre), and the Text Encoder extracts semantic and syntactic features from the corrected text. These embeddings are fused by a synthesizer to generate stutter-free audio.

\tool{} adopts a zero-shot, diffusion-based voice cloning model, leveraging StyleTTS2~\cite{StyleTTS} for high-fidelity and speaker-consistent synthesis, even from limited audio samples. To reduce latency, \tool{} processes speech segment by segment. Repair begins once the initial portion of input is received and outputs are generated shortly after each segment ends, eliminating the delay of waiting for full utterances.

\subsection{Implementation}
We developed and deployed \tool{} as a web application with a custom-built user interface and a Python Flask backend. The machine learning models were implemented using PyTorch and Transformers. Key components include Whisper-base for speech-to-text conversion, GPT-4o for semantic-aware text repair, and StyleTTS2 for text-to-speech conversion. During the user study, \tool{} ran on a server with an Intel Xeon Gold 6430 (16 cores), 120GB RAM, and an RTX 4090 (24GB) GPU.

\begin{table*}[t]
	\begin{center}
        \begin{tabular}{ccccccc}
            \hline 
            \textbf{Models} & \multicolumn{2}{c}{\textbf{WER (\%)}} & \multicolumn{2}{c}{\textbf{MER (\%)}} & \multicolumn{2}{c}{\textbf{WIL (\%)}} \\
            \cline{2-7}
             & Original & \tool & Original & \tool & Original & \tool \\
            \hline             
            data2vec  & 27.06 & 14.63 & 21.79 & 13.05 & 28.58 & 17.66 \\
            wav2vec2  & 35.93 & 23.11 & 35.13 & 22.23 & 49.04 & 30.15 \\
            whisper   & 21.53 & 15.38 & 18.74 & 14.68 & 24.87 & 20.48 \\
            SenseVoice(Mandarin)   & 41.86 & 21.65 & 34.02 & 20.40 & 39.43 & 22.89 \\
            \hline
            Average  & 31.60 & 18.69 & 27.42 & 17.59 & 35.48 & 22.80 \\
            \hline  
        \end{tabular}
	\end{center}
		\caption{Comparative Analysis of ASR System Performance on Stuttering Audio Repair. Metrics include Word Error Rate (WER), Match Error Rate (MER), and Word Information Lost (WIL) for original and \tool{}-processed speech. “data2vec” stands for \texttt{data2vec-audio-large-960h}, “wav2vec2” for \texttt{wav2vec2-large-xlsr-53-english}, “whisper” for \texttt{whisper-base}, and "SenseVoice" for \texttt{SenseVoice-small}}
		\label{effect}
\end{table*}

\section{Evaluation}

\subsection{Experimental Setup}
To evaluate \tool{}’s effectiveness, we used the English subset of FluencyBank~\cite{fluency-bank}, containing 25 hours of stuttering speech from 48 speakers (21 female, 27 male; aged 10–70). Of 245 recordings, 128 child–adult dialogues with short or fragmented utterances were excluded as unsuitable for evaluation. To assess cross-lingual adaptability, \tool{} was also tested on the AS-70 Mandarin dataset~\cite{gong24} with 70 hours of stuttering speech. Five rounds of testing were conducted using three open-source ASR models: Whisper-base, data2vec-audio, and wav2vec2. These models were selected for their efficiency, stability, and widespread use. For Mandarin data, we employed SenseVoice, noted for its robust Mandarin recognition performance. System performance was evaluated using four standard metrics:

     \textbf{(1) Word Error Rate (WER)}: Calculates the proportion of word-level insertions, deletions, and substitutions in ASR output compared to the reference.
     
     \textbf{(2) Match Error Rate (MER)}: Measures how well the predicted and reference transcripts align, including positional and sequencing accuracy.
     
     \textbf{(3) Word Information Lost (WIL)}: Estimates the proportion of linguistic content lost during ASR transcription.
     
     \textbf{(4) Semantic Similarity}: Quantifies the semantic alignment between repaired text and ground truth using Sentence-BERT~\cite{Sentence-BERT}.

\begin{table}[t]
\centering
\begin{tabular}{ccc}
    \hline 
    \textbf{Models} & \textbf{Original} & \textbf{Repair} \\
    \hline             
    data2vec  & 0.74 & 0.90 \\
    wav2vec2  & 0.85 & 0.93 \\
    whisper  & 0.83 & 0.89 \\
    \hline  
    Average  & 0.81 & 0.91 \\
    \hline  
\end{tabular}
\caption{Comparison of Semantic Similarity Before and After Repairs Across Different ASR Models}
\label{semantic}
\end{table}

\begin{table}[t]
        \centering
        \begin{tabular}{cccc}
        \hline
        \textbf{Models} & \textbf{WER(\%)} & \textbf{MER(\%)} & \textbf{WIL(\%)} \tabularnewline
        \hline
        data2vec  & 19.19 & 18.85 &25.67  \tabularnewline
        wav2vec2 & 23.52 & 22.12 &30.90 \tabularnewline
        whisper               & 16.50 & 15.73 & 21.82   \tabularnewline
        \hline
        Average               & 19.74 & 18.90 & 26.13   \tabularnewline
        \hline
        \end{tabular}
        \caption{Performance Metrics Including Word Error Rate (WER), Match Error Rate (MER), and Word Information Lost (WIL) for Clone Audio Error Rate Analysis}
        \label{Clone Audio}
\end{table}

\begin{table}[t]
        \centering
        \begin{tabular}{cccc}
        \hline
        \textbf{Model}            & \textbf{WER(\%)}   & \textbf{MER(\%)}   & \textbf{WIL(\%)}   \\ \hline
        whisper(original)        & 21.53     & 18.74      & 24.87       \\ \hline
        \textbf{whisper(1e-5)} & \textbf{18.32} & \textbf{16.09} & \textbf{21.98} \\ \hline
        whisper(1e-6)         & 20.86        & 17.89        & 24.34        \\ \hline
        \end{tabular}
        \caption{Performance comparison of Whisper-base models fine-tuned with different learning rates} 
        \label{Fine tuning}
\end{table}

\begin{table}[t]
\centering
\setlength{\tabcolsep}{1mm}
\begin{tabular}{ccccccc}
    \hline
    \multirow{2}{*}{Severity} & \multicolumn{3}{c}{\textbf{Original (\%)}} & \multicolumn{3}{c}{\textbf{Repaired (\%)}} \\
    \cline{2-7}
    & WER & MER & WIL & WER & MER & WIL \\
    \hline
    Severe   & 33.48 & 31.04 & 40.55 & 22.62 & 22.28 & 31.29 \\
    Moderate & 23.46 & 22.47 & 34.29 & 16.88 & 16.46 & 25.03 \\
    Mild     & 14.16 & 13.89 & 20.42 & 11.01 & 10.96 & 15.56 \\
    \hline  
    Average  & 23.70 & 22.47 & 31.75 & 16.84 & 16.57 & 23.96 \\
    \hline
\end{tabular}
\caption{Comparative Analysis of \tool{} Performance for Different Stuttering Severity Levels}
\label{severity_comparison}
\end{table}

\subsection{Result and Analysis}

As shown in Table~\ref{effect}, \tool{} significantly reduces ASR error rates by providing repaired, stuttering-free audio. Post-processing with \tool{} leads to substantial improvements in WER, MER, and WIL across various ASR systems, demonstrating its effectiveness in mitigating the impact of stuttering on speech recognition accuracy.

To assess semantic improvements introduced by \tool{}, we measured the similarity between both the original ASR transcriptions and \tool{}'s repaired texts against ground truth. As shown in Table~\ref{semantic}, scores improved from 0.74–0.85 (ASR) to 0.89–0.93 (repaired), indicating stronger semantic fidelity. Additionally, we reprocessed the repaired audio through the original ASR systems. As shown in Table~\ref{Clone Audio}, error rates were consistent with those from text repairs, confirming that \tool{} maintains both transcription accuracy and semantic intent, effectively improving communication for PWS.

\subsection{Ablation} 
We conducted ablation studies comparing zero-shot prompting, removing the multi-agent framework, and varying the number of repair agents. Each setup was tested five times using three ASR models with consistent metrics.

The results reveal three findings. First, \tool{} significantly outperforms the zero-shot prompting approach, demonstrating the effectiveness of our prompt engineering strategy, which incorporates CoT reasoning. Second, removing the multi-agent framework leads to a notable drop in performance, confirming the importance of modular agent collaboration. Third, increasing the number of repair agents improves performance initially, but gains plateau beyond three agents while latency continues to rise. Based on these observations, we adopt a three-agent configuration to strike a balance between correction quality and system responsiveness. Full results are provided in the supplementary material.

\subsection{Real-world Application}

This section explores the practical effectiveness of \tool{}, which generates both repaired textual transcripts and stuttering-free audio. We examine its application in three real-world scenarios:

\textbf{Stuttering-free Audio as a Reference.} The FluencyBank dataset primarily contains stuttering audio samples. By incorporating \tool{}'s repaired speech, we enhance the dataset with stuttering-free references, providing valuable comparative data for linguistic and therapeutic research. Our collaboration with the FluencyBank team validated the utility of \tool{} in enhancing dataset quality, and highlighted its potential for supporting stuttering diagnosis and therapeutic research.

\textbf{Stuttering Audio Annotation.} Many stuttering datasets lack comprehensive textual annotations, making \tool{} essential for automatic annotation. We applied \tool{} to the SEP-28K dataset~\cite{Lea2021SEP28kAD}, which lacks ground truth annotations. \tool{} provided accurate text repairs, facilitating a collaborative annotation effort and demonstrating its capability to streamline data preparation for speech disorder research.

\textbf{Fine-tuning ASR Models.} Fine-tuning ASR models with \tool{}’s repaired audio improved recognition accuracy for stuttered speech. As shown in Table~\ref{Fine tuning}, the fine-tuned models achieved consistent gains across metrics, confirming \tool{}’s effectiveness in enhancing ASR adaptability to speech disorders.

\section{User Study}

We conducted two user studies to evaluate \tool{}’s usability and effectiveness. The Closed-Ended Task Study measured repair accuracy in controlled reading tasks, while the Open-Ended Task Study examined real-time usability in natural interactions, demonstrating \tool{}’s practical value. A total of 23 participants (ages 18–45; 15 males, 8 females) with over seven years of stuttering experience were recruited under institutional ethical approval. Stuttering severity, assessed using the Andrew \& Harris Scale~\cite{AH}, was classified as mild (7), moderate (12), and severe (4). A subsequent perceptual evaluation with 12 PWS confirmed that the perceived fluency gains stemmed from genuine system performance rather than placebo effects.

\subsection{Close-Ended Task Study}

\textbf{Task.} Each participant was randomly assigned one of three standardized 150-word English passages to read aloud. Their speech recordings were processed by \tool{} for both semantic-aware text repair and voice synthesis. Participants then rated the original and repaired outputs (both text and audio) using 5-point Likert scales. Internal consistency was validated via Cronbach's $\alpha$~($> 0.7$), and we report the results as means with 95\% confidence intervals.

\textbf{Performance.} Participants reported high satisfaction with the system-generated text ($\text{mean} = 4.30 \pm 0.27$, $p \ll 0.01$). To further evaluate the repaired output, we conducted a two-fold analysis: (1) textual quality in terms of grammar, phrasing, and readability; and (2) repair effectiveness, focusing on the correction of stuttering while preserving the original semantics.

Participants were asked to identify issues such as grammatical errors or unnatural phrasing in the repaired text. Results showed that they generally found the output to be natural and fluent ($\text{mean} = 4.13 \pm 0.35$, $p \ll 0.01$) and required minimal further editing ($\text{mean} = 3.74 \pm 0.39$, $p \ll 0.01$). This indicates that \tool{} can generate high-quality, readable text with little need for post-processing.

In another task, participants compared the ASR transcription and the version repaired by \tool{} while listening to the original stuttered audio. They agreed that the system effectively corrected disfluencies ($\text{mean} = 4.35 \pm 0.30$, $p \ll 0.01$) and largely preserved the original meaning ($\text{mean} = 4.13 \pm 0.35$, $p \ll 0.01$). Additionally, results from Table~\ref{severity_comparison} shows that \tool{} reduces WER from 23.10\% to 16.17\%, MER from 21.99\% to 15.87\%, and WIL from 31.83\% to 23.35\%. While improvements are more significant in moderate and severe stuttering, even mild stuttering shows benefits-particularly in reducing WIL. These results confirm the practical applicability of \tool{} in enhancing ASR accuracy across varying stuttering severities.

Participants noted that the transcribed text could be utilized in important situations, such as interviews or meetings, to accurately record their speech and help users understand how their speech would sound without stuttering. Since \tool{} can produce transcripts free of stuttering, it enables users to record their intended messages more accurately and conveniently, without the interference of disfluencies. One participant suggested that this feature could be integrated into meeting software's auto-generated text functions, allowing for higher accuracy in the transcription of speech from PWS during meetings.

\subsection{Open-Ended Task Study}

We conducted a within-subjects user study with the same 23 participants to assess \tool{}'s usability in open-ended tasks. The protocol closely followed Study 1, except participants listened to recordings of real-time conversations with researchers instead of reading fixed passages in a questionnaire. Unlike Study 1's scripted reading, Study 2 involved 7-10 minute one-on-one online conversations between participants and a researcher. Topics focused on stuttering's impact and coping strategies, with participants speaking naturally while \tool{} performed real-time speech repair. We evaluated \tool{}'s real-time performance across three aspects:

     \textbf{(1) User Satisfaction and Usability Evaluation.} We used Net Promoter Score (NPS) to assess users' willingness to recommend \tool{}, and  System Usability Scale (SUS) to evaluate perceived ease of use and learnability.

     \textbf{(2) Execution Time.} Real-time latency was measured from input to output to evaluate system responsiveness.

    \textbf{(3) Synthesized Speech Quality.} We evaluated the repaired audio using Mean Opinion Score (MOS) ratings on naturalness and similarity to assess voice consistency.

\textbf{Performance.} We calculated a Cronbach’s $\alpha$ of 0.73 for the questionnaire, indicating acceptable reliability. Regarding the system's overall effectiveness, most participants expressed positive opinions ($mean = 4.22 \pm 0.36(p \ll 0.01)$) and believed that \tool{} could effectively assist in addressing communication issues related to stuttering ($mean = 3.91 \pm 0.34(p \ll 0.01)$).

Participants reported high satisfaction with the quality of the repaired audio ($mean = 4.39 \pm 0.25(p \ll 0.01)$), indicating that the system effectively produced polished and cohesive audio. The score of \tool{} on MOS-N is $4.52 \pm 0.35(p \ll 0.01)$, indicating that participants consider the repaired audio to be smooth and coherent, and tool {} successfully reduces interruptions or unnatural pauses.

When comparing the repaired audio to the original, participants widely agreed that \tool{} significantly reduced stuttering ($mean = 4.70 \pm 0.49(p \ll 0.01)$), highlighting the system's effectiveness in addressing disfluent and halting speech. However, feedback on the similarity between the synthetic and original audio was mixed. The rating of \tool {} on MOS-S is $3.4 \pm 0.36 (p \ll 0.5)$. Participants noted only a partial resemblance between the two versions, attributing this to the challenges of perfectly replicating vocal nuances and the use of a zero-shot cloning approach. The variability in similarity ratings stems from a trade-off between repair speed, quality, and generalizability. We opted for zero-shot cloning to enhance generalizability, allowing \tool{} to work with any speaker without prior voice samples. This approach prioritizes processing speed to minimize latency, albeit at the expense of some voice naturalness. 

\tool{} achieved an SUS score of 66.9, indicating acceptable usability, and an NPS of 23.53, suggesting a generally favorable user attitude. 

\textbf{User Feedback.} Open-ended responses and follow-up interviews revealed generally positive impressions of \tool{}. Several participants noted that the repaired audio conveyed meaning more clearly and restored natural speech flow. One commented, ``\tool{} helps me express myself more fluently and be better understood." Many saw potential use in formal settings such as public speaking and meetings. Some concerns were raised about voice naturalness. Seven participants felt the synthesized voice lacked warmth and did not fully resemble the original speaker. This is likely due to our zero-shot cloning approach, which prioritizes generalizability over high-fidelity reproduction.

We also measured system latency, with an average of $3.77 \pm 0.62$ seconds per ~50-100-word segment. Most participants found this acceptable for longer speech contexts ($mean = 4.08 \pm 0.32$), though two mentioned it may be disruptive in spontaneous conversations. These insights suggest areas for further optimization in voice quality and processing speed to support real-time use.

\section{Discussion}

\noindent \textbf{Linguistic and Paralinguistic Integrity Issues.} By removing fillers and repetitions, \tool{} may alter paralinguistic cues (tone, pitch, and rhythm), potentially affecting conversational nuance. While these elements can signal emphasis or hesitation, they often obscure meaning in stuttered speech. Given that listeners typically cannot distinguish intentional repetition from disfluency, their removal represents a reasonable trade-off that improves clarity and intelligibility. Similar disfluency removal attempts by ASR systems like Whisper remain limited, whereas \tool{} achieves greater fluency while preserving core semantics.

\noindent\textbf{Data and System Coverage Constraints.} 
Our evaluation focused on mainstream ASR systems (Google, Azure, OpenAI) chosen for their popularity and recent updates. While this introduces potential selection bias, it addresses the primary barriers PWS face with current voice technology, and our consistent improvements across these diverse architectures suggest broader applicability. Additionally, we tested on both English and Chinese speech due to the limited availability of stuttering datasets, highlighting the data scarcity that contributes to PWS exclusion from AI systems. While further validation across accents and languages is needed, our results demonstrate the feasibility and extensibility of the multi-agent repair framework.

\section{Conclusion and Future Work}

In this paper, we present \tool{}, a novel framework that converts  stuttering speech into fluent audio. The process begins with converting spoken words into text via advanced ASR technologies. Subsequently, an ensemble of AI agents corrects the transcription, and the refined text is then re-converted into spoken words using the Styletts voice cloning system, preserving the speaker's authentic voice characteristics. This end-to-end solution serves a dual purpose: it provides seamless audio for PWS, and it supplies developers with enhanced tools for the better recognition and annotation of stuttered speech. Our future work target further refining the AI's ability to detect and rectify stuttered speech more precisely, broadening the framework's applicability to a wider range of speech anomalies.

\section*{Acknowledgments}
This work is supported by National Key Research and Development Program 
(No. 2022YFB3305200), National Natural Science Foundation of China (No. 62441605), the ``Digital Silk Road'' Shanghai International Joint Lab of Trustworthy Intelligent Software (Grant No. 22510750100), and Shanghai Trusted Industry Internet Software Collaborative Innovation Center.
\bibliography{aaai2026}

\end{document}